\documentclass[twocolumn,prl,floatfix,superscriptaddress,nobibnotes]{revtex4-1}

\usepackage{amsmath}
\usepackage{amssymb}
\usepackage{graphicx}
\usepackage{bm}
\usepackage{color}
\usepackage[separate-uncertainty=true,alsoload=torr]{siunitx}

\begin{document}

\title{Selective Area Grown Semiconductor-Superconductor Hybrids:\\ A Basis for Topological Networks}

\author{S.~Vaitiek\.{e}nas}
\affiliation{Center for Quantum Devices and Station Q Copenhagen, Niels Bohr Institute, University of Copenhagen, Copenhagen, Denmark}
\author{A.~M.~Whiticar}
\affiliation{Center for Quantum Devices and Station Q Copenhagen, Niels Bohr Institute, University of Copenhagen, Copenhagen, Denmark}
\author{M.-T.~Deng}
\affiliation{Center for Quantum Devices and Station Q Copenhagen, Niels Bohr Institute, University of Copenhagen, Copenhagen, Denmark}
\author{F.~Krizek}
\affiliation{Center for Quantum Devices and Station Q Copenhagen, Niels Bohr Institute, University of Copenhagen, Copenhagen, Denmark}
\author{J.~E.~Sestoft}
\affiliation{Center for Quantum Devices and Station Q Copenhagen, Niels Bohr Institute, University of Copenhagen, Copenhagen, Denmark}
\author{C.~J.~Palmstr\o{}m}
\affiliation{Materials Department, University of California, Santa Barbara, CA 93106, USA}
\author{S.~Marti-Sanchez}
\affiliation{Catalan Institute of Nanoscience and Nanotechnology (ICN2), CSIC and BIST, Campus UAB, Bellaterra, Barcelona, Catalonia, Spain}
\author{J.~Arbiol}
\affiliation{Catalan Institute of Nanoscience and Nanotechnology (ICN2), CSIC and BIST, Campus UAB, Bellaterra, Barcelona, Catalonia, Spain}
\affiliation{ICREA, Pg. Llu\'is Companys 23, 08010 Barcelona, Catalonia, Spain}
\author{P.~Krogstrup}
\affiliation{Center for Quantum Devices and Station Q Copenhagen, Niels Bohr Institute, University of Copenhagen, Copenhagen, Denmark}
\author{L.~Casparis}
\affiliation{Center for Quantum Devices and Station Q Copenhagen, Niels Bohr Institute, University of Copenhagen, Copenhagen, Denmark}
\author{C.~M.~Marcus}
\affiliation{Center for Quantum Devices and Station Q Copenhagen, Niels Bohr Institute, University of Copenhagen, Copenhagen, Denmark}

\date{\today}

\begin{abstract}
We introduce selective area grown hybrid InAs/Al nanowires based on molecular beam epitaxy, allowing arbitrary semiconductor-superconductor networks containing loops and branches. Transport reveals a hard induced gap and unpoisoned 2{\em e}-periodic Coulomb blockade, with temperature dependent 1{\em e} features in agreement with theory. Coulomb peak spacing in parallel magnetic field displays overshoot, indicating an oscillating discrete near-zero subgap state consistent with device length. Finally, we investigate a loop network, finding strong spin-orbit coupling and a coherence length of several microns. These results demonstrate the potential of this platform for scalable topological networks among other applications.

\end{abstract}

\maketitle
Majorana zero modes (MZMs) at the ends of one-dimensional topological superconductors are expected to exhibit non-Abelian braiding statistics \cite{Kitaev2000,Read2000}, providing naturally fault-tolerant qubits \cite{Nayak2008, DasSarma2015}. Proposed realizations of braiding \cite{Alicea2011,Aasen2016}, interference-based topological qubits \cite{Flensberg2011,Plugge2017,Vijay2016} and topological quantum computing architectures \cite{Karzig2017} require scalable nanowire networks. While relatively simple branched or looped wires can be realized by specialized growth methods \cite{Gazibegovic2017,Krogstrup2015} or by etch- and gate-confined two-dimensional hybrid heterostructures \cite{Shabani2016, Kjaergaard2016, Suominen2017, Nichele2017}, selective area growth \cite{Krizek2018} enables deterministic patterning of arbitrarily complex structures. This allows complex continuous patterns of superconductor-semiconductor hybrids and topological networks.

Following initial theoretical proposals \cite{Lutchyn2010,Oreg2010}, a number of experiments have reported signatures of Majorana zero modes (MZMs) in hybrid semiconductor-superconductor nanowires \cite{Krogstrup2015}, including zero-bias conductance peaks \cite{Mourik2012,Das2012,Gul2018,Deng2016, Suominen2017, Nichele2017,Zhang2017,Deng2012,Sestoft2017} and Coulomb blockade peak spacing oscillations \cite{Albrecht2016,Sherman2016}. To date, experiments have used individual vapor-liquid-solid (VLS) nanowires \cite{Mourik2012,Das2012,Gul2018,Deng2016,Zhang2017,Deng2012} or gate-confined two-dimensional heterostructures \cite{Suominen2017,Nichele2017}. Within these approaches, constructing complex topological devices and networks containing branches and loops \cite{Alicea2011,Aasen2016,Flensberg2011,Vijay2016,Karzig2017,Plugge2017} is a challenge. Recently, branched and looped VLS growth has been developed toward this goal \cite{Krizek2017,Gazibegovic2017}. 

\begin{figure}[b]
\includegraphics[width=\linewidth]{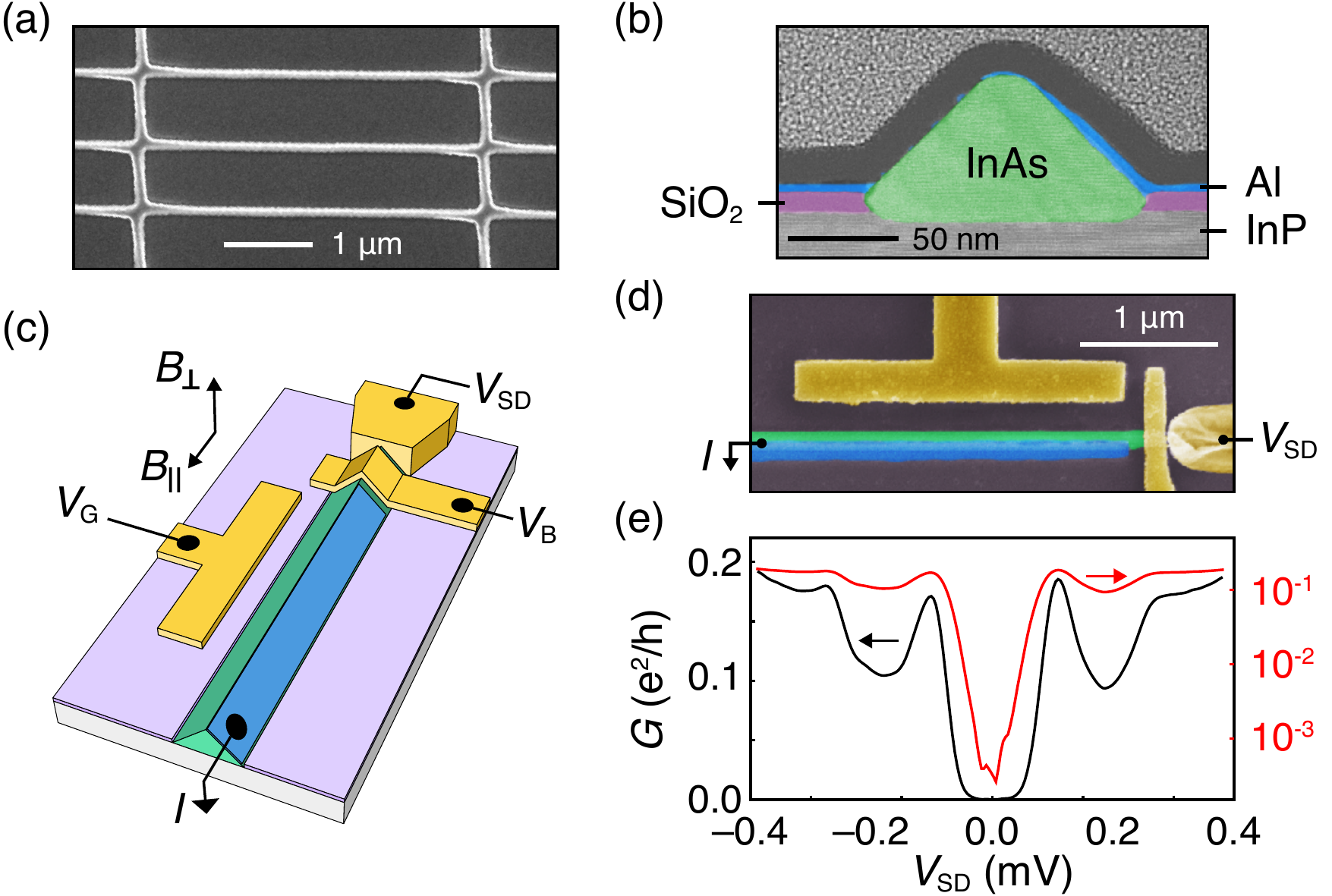}
\caption{\label{fig:1} (a) Scanning electron micrograph of a SAG hybrid network. (b) False-colored annular dark field scanning transmission electron micrograph of a nanowire cross-section displays InP substrate, InAs (green) nanowire, Al (blue) shell and SiO$_2$ (purple) mask. (c) Measurement set-up for device 1 showing the gate voltages and orientations of magnetic fields used in the measurements. (d) False-colored electron micrograph of device 1. (e) Differential conductance, $G$, as a function of source-drain bias, $V_{\rm SD}$, in linear (black) and logarithmic (red) scales shows a hard superconducting gap. 
}
\end{figure}

In this Letter, we investigate a novel approach to the growth of semiconductor-superconductor hybrids that allows deterministic on-chip patterning of topological superconducting networks based on SAG. We characterize key physical properties required for building Majorana networks, including a hard superconducting gap, induced in the semiconductor, phase-coherence length of several microns, strong spin-orbit coupling, and Coulomb blockade peak motion compatible with interacting Majoranas. Overall, these properties show great promise for SAG-based topological networks.

\begin{figure}[t]
\includegraphics[width=\linewidth]{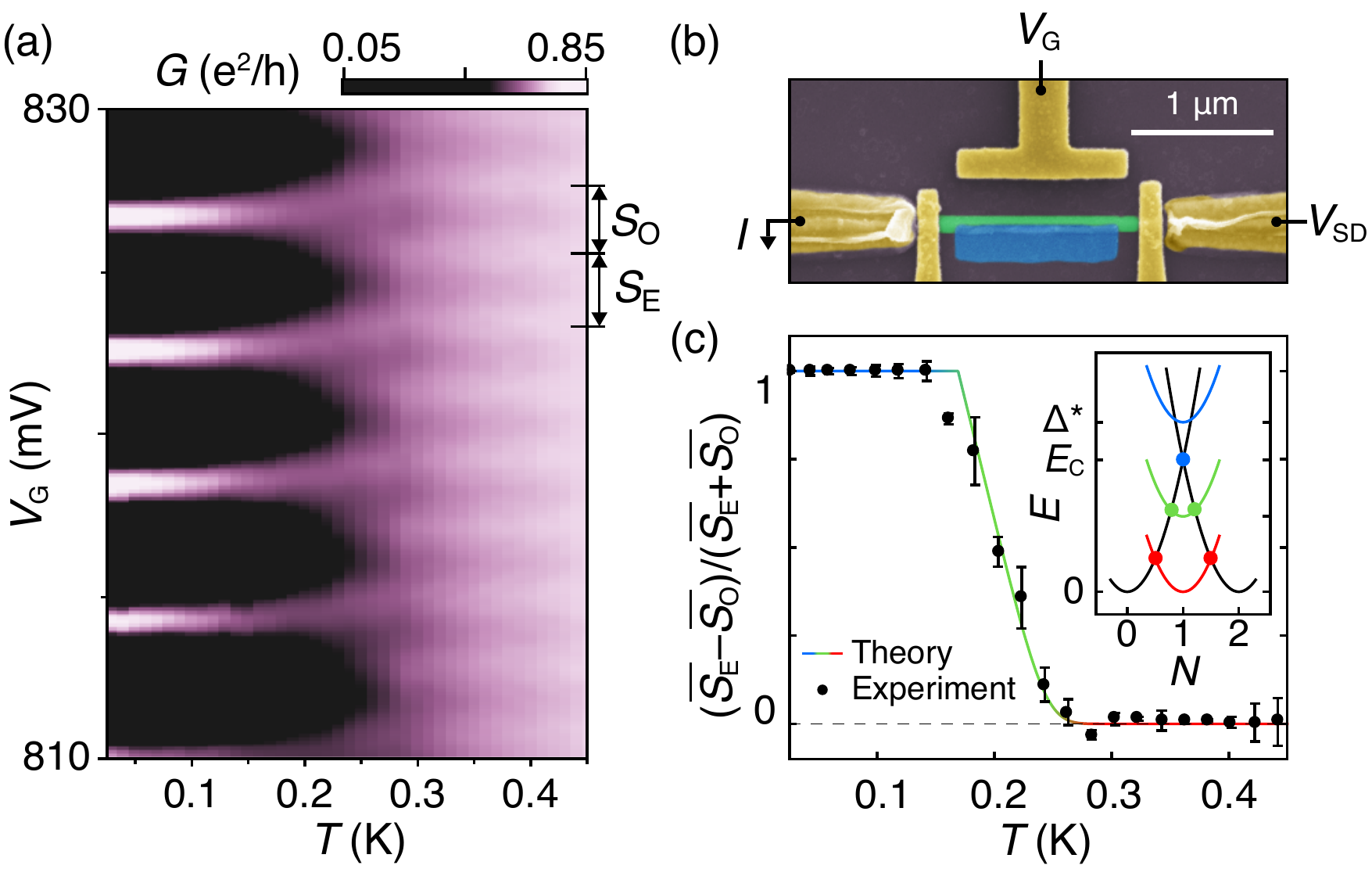}
\caption{\label{fig:2} (a) Conductance, $G$, of Device 2 as a function of applied gate voltage, $V_{\rm G}$, and temperature, $T$. A characteristic 2{\em e} to 1{\em e} transition occurs around $\SI{200}{\milli\kelvin}$. The color scale was adjusted for better visibility. (b) False-colored electron micrograph of Device 2. (c) Normalized even-odd peak spacing difference, $(\overline{S_{\rm E}}-\overline{S_{\rm O}})/(\overline{S_{\rm E}}+\overline{S_{\rm O}})$, from the measurements shown in (b) as a function of $T$.  The error-bars were estimated using standard deviation of the peak spacing. The theoretical fit corresponds to an induced superconducting gap $\Delta^*=\SI{190}{\mu\electronvolt}$. Inset: Energy, $E$, of the device as a function of normalized gate voltage, $N$. Black (colour) parabolas correspond to even(odd)-parity ground state. Transport occurs at the charge degeneracy points indicated by filled circles.
}
\end{figure}

Selective area growth was realized on a semi-insulating $\left(001\right)$ InP substrate. PECVD grown SiO$_{\rm x}$ was patterned using electron beam lithography and wet etching. InAs wires with triangular cross-sections were grown by molecular beam epitaxy (MBE). The Al was grown {\it in-situ} by MBE using angled deposition covering one of the facets. The excess Al was removed by wet etching [Fig.~\ref{fig:1}(a-c)]. The details of the semiconductor growth are given in Ref.~\cite{Krizek2018}, while it is superconductor-semiconductor proximity effects that are emphasized in the present study. Data from four devices are presented. Device 1 [Fig.~\ref{fig:1}(d)] consists of a single barrier at the end of a $\SI{4}{\mu\meter}$ wire, defined by a lithographically patterned gate adjacent to a Ti/Au contact where the Al has been removed by wet etching. This device allowed density of states measurement at the end of the wire by means of bias spectroscopy, to investigate the superconducting proximity effect in the InAs. Evolution of Coulomb blockade in temperature and magnetic field was studied in Device 2 [Fig.~\ref{fig:2}(b)]---a hybrid quantum dot with length of $\SI{1.1}{\mu\meter}$ defined by two Ti/Au gates adjacent to etched-Al regions. The barrier voltages $V_{\rm B}$ were used to create tunneling barriers. The chemical-potential in the wires was tuned with gate voltage $V_{\rm G}$. Device 3 was a micron-size square loop [Fig.~\ref{fig:4}(a)] with fully removed Al, which was used to extract phase coherence lengths from weak antilocalization (WAL) and Aharonov-Bohm (AB) oscillations. Device 4 [Fig.~S~2 in the Supplemental Material \cite{SupMaterial}]---top-gated nanowire without Al---was used to extract the charge carrier mobility. The nanowires in Devices 1, 2 and 4 are parallel to $[100]$ direction, whereas the arms of Device 3 are oriented along $[100]$ and $[010]$ directions. Standard ac lock-in measurements were carried out in a dilution refrigerator with a three-axis vector magnet. See Supplemental Material \cite{SupMaterial} for more detailed description of the growth, fabrication and measurement setup.

Differential conductance, $G$, in the tunneling regime, as a function of source-drain bias, $V_{\rm SD}$, for Device 1 [Fig.~\ref{fig:1}(e)] at $V_{\rm G}=\SI{-9.2}{\volt}$ reveals a gapped density of states with two peaks at $V_{\rm SD}=\SI{110}{\mu\electronvolt}$ and $\SI{280}{\mu\electronvolt}$. We tentatively identify the two peaks with two populations of carriers in the semiconductor, the one with a larger gap residing at the InAs-Al interface and with a smaller at the InAs-InP. The magnitude of the larger superconducting gap is consistent with enhanced energy gaps of $\SI{290}{\mu\electronvolt}$ for $\SI{7}{\nano\meter}$ Al film \cite{Court2008}. The zero-bias conductance is $\sim 400$ times lower than the above-gap conductance, a ratio exceeding VLS nanowire \cite{Chang2015,Gazibegovic2017,Zhang2017_2} and 2DEG devices \cite{Kjaergaard2016}, indicating a hard induced gap. We note, however, that co-tunneling through a quantum-dot or multichannel tunneling can enhance this ratio \cite{Beenakker1992}. The spectrum evolution with $V_{\rm G}$ from enhanced to suppressed conductance around $V_{\rm SD}=\SI{0}{\milli\volt}$ is shown in Fig.~S~1 in Supplemental Material \cite{SupMaterial}.

Transport through a Coulomb island geometry [Fig.~\ref{fig:2}] at low temperatures shows 2{\em e}-periodic peak spacing as a function of $V_{\rm G}$. Coulomb diamonds at finite bias yield a charging energy $E_{\rm C}=\SI{60}{\mu\electronvolt}$ (see Fig.~S1 in Supplemental Material \cite{SupMaterial}), smaller than the induced gap, $\Delta^{*}\,\sim\,\SI{100}{\mu\electronvolt}$, as seen in Fig.~\ref{fig:1}(e). The zero-bias Coulomb blockade spacing evolves to even-odd and finally to 1{\em e}-periodic peaks with increasing temperature,~$T$. The 2{\em e} to 1{\em e} transition in temperature does not result from the destruction of superconductivity, but rather arises due to the thermal excitation of quasiparticles on the island, as investigated previously in metallic islands \cite{Tuominen1992,Lafarge1993} and semiconductor-superconductor VLS nanowires \cite{Higginbotham2015}.

A thermodynamic analysis of Coulomb blockade peak spacings is based on the difference in free energies, $F= F_{\rm O} - F_{\rm E}$, between even and odd occupied states. We consider a simple model that assumes a single induced gap $\Delta^{*}$, not accounting for the double-peaked density of states in Fig.~\ref{fig:1}(e). At low temperatures ($T \ll E_{\rm C},\Delta^{*}$), $F$ approaches $\Delta^{*}$. Above a characteristic poisoning temperature, $T_{\rm p}$, quasiparticles become thermally activated and $F$  decreases rapidly to zero. For $F(T) > E_{\rm C}$, Coulomb peaks are 2{\em e} periodic with even peak spacings, $S_{\rm E}\propto E_{\rm C}$, independent of $T$. For $F(T) < E_{\rm C}$, odd states become occupied, and the difference in peak spacing, $S_{\rm E} - S_{\rm O}$, decreases roughly proportional to $F$. A full analysis following Ref.~\cite{Higginbotham2015} (see Supplemental Material \cite{SupMaterial}) yields the peak spacing difference
\begin{equation}\label{eq:1}
\begin{split}
S_{\rm E}-S_{\rm O} & = \frac{2}{\eta e} \min\left(E_{\rm C},F \right) \\
& = (S_{\rm E}+S_{\rm O}) \min\left(1,F/E_{\rm C} \right),
\end{split}
\end{equation}
where $\eta$ is the dimensionless gate lever arm measured from Coulomb diamonds.

Figure~\ref{fig:2}(c) shows the measured even-odd difference in peak spacing, $(\overline{S_{\rm E}}-\overline{S_{\rm O}})/(\overline{S_{\rm E}}+\overline{S_{\rm O}})$, averaged over 4 peaks in Device 2, along with Eq.~(\ref{eq:1}). Thermodynamic analysis shows an excellent agreement with the peak spacing data across the full range of temperatures. The fit uses an independently measured $E_{\rm C}$, with the induced gap as a single fit parameter, yielding $\Delta^*=\SI{190}{\mu\electronvolt}$, a reasonable value that lies between the two density of states features in Fig.~\ref{fig:1}(e). The island remains unpoisoned below $T_{\rm p}~\sim \SI{250}{\milli\kelvin}$.

\begin{figure}[t]
\includegraphics[width=\linewidth]{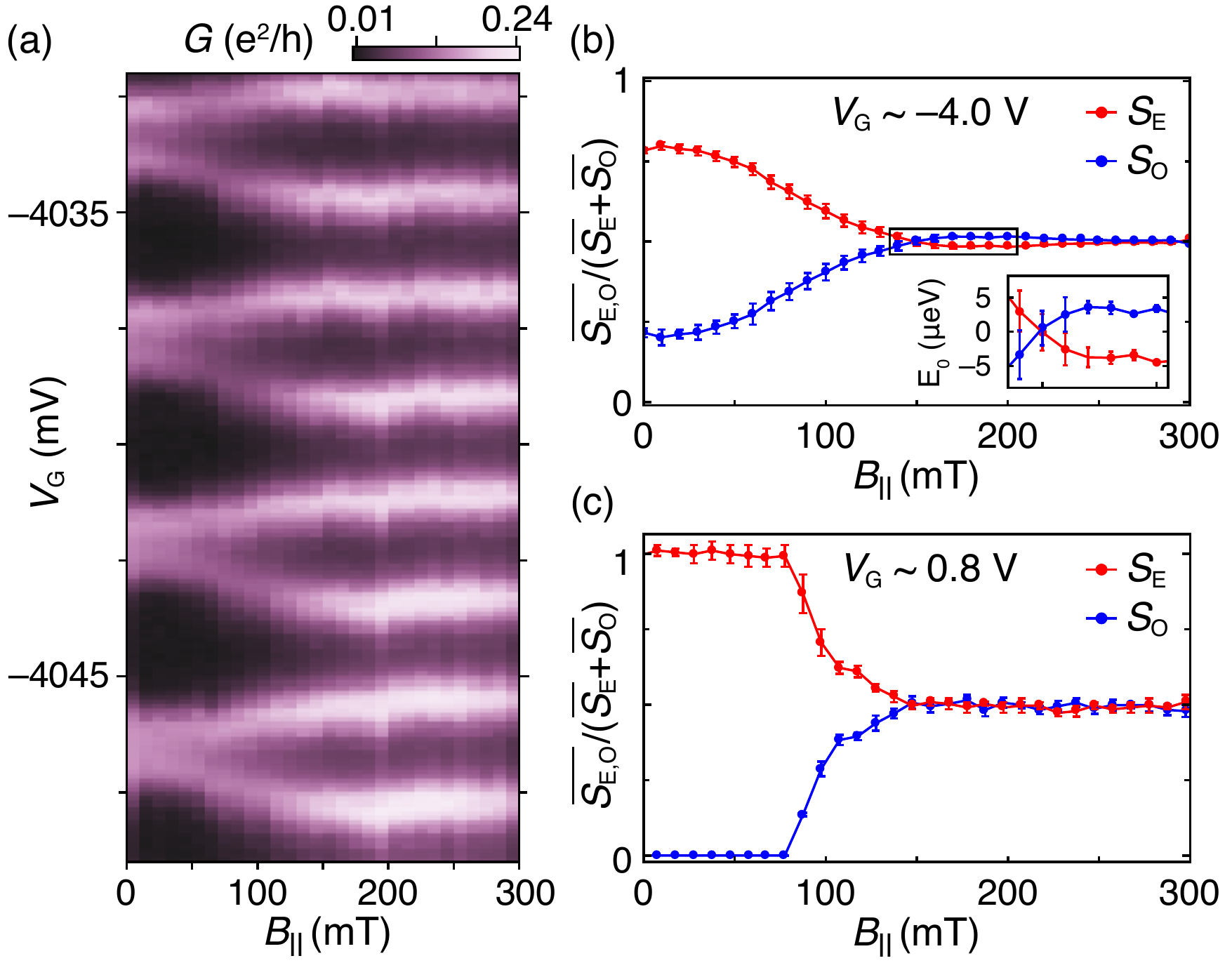}
\caption{\label{fig:3} (a) Conductance as a function of $V_{\rm G}$ and $B_{\parallel}$ from Device 2, taken at $V_{\rm G}\sim\SI{-4.0}{\volt}$, shows even-odd peak spacings at zero field transit to 1{\em e} spacing when the field is increased. (b) Normalized even and odd Coulomb peak spacings, $\overline{S_{\rm E,O}}/(\overline{S_{\rm E}}+\overline{S_{\rm O}})$, from the measurements shown in \textbf{a}, as a function of $B_{\parallel}$. Inset: zoom-in of the peak spacing overshoot with amplitude of $\SI{7}{\mu\electronvolt}$ at $B_{\parallel}=\SI{170}{\milli\tesla}$. (c) Same as (b), but taken at $V_{\rm G}\sim\SI{0.8}{\volt}$. At positive gate voltage, the peaks become evenly spaced above $B_{\parallel}=\SI{150}{\milli\tesla}$. The error-bars in (a) and (b) were estimated using the standard deviation of the peak position.
}
\end{figure}

The evolution of Coulomb blockade peaks with parallel magnetic field, $B_\parallel$, is shown in Fig.~\ref{fig:3}(a). In this data set, peaks show even-odd periodicity at zero field due to a gate-dependent gap or a bound state at energy $E_0$ less than $E_{\rm C}$. A subgap state results in even-state spacing proportional to $E_{\rm C}+E_0$ and odd-state spacing $E_{\rm C}-E_0$ \cite{Albrecht2016} (see Supplemental Material \cite{SupMaterial}), giving
\begin{equation}\label{eq:2}
\begin{split}
S_{\rm E,O} & = \frac{1}{\eta e} \left[E_{\rm C} \pm \min\left(E_{\rm C},E_0 \right) \right] \\
& = \frac{S_{\rm E}+S_{\rm O}}{2} \left[1 \pm \min\left(1,E_0/E_{\rm C}\right) \right].
\end{split}
\end{equation}

Figure \ref{fig:3}(b) shows the $B_\parallel$ dependence of even and odd peak spacings, $\overline{S_{\rm E,O}}/(\overline{S_{\rm E}}+\overline{S_{\rm O}})$, extracted from the data in Fig. \ref{fig:3}(a), giving an effective $g$-factor of $\sim 13$. Even and odd peak spacings become equal at $B_\parallel=\SI{150}{\milli\tesla}$, then overshoot at higher fields with a maximum amplitude corresponding to $\SI{7(1)}{\mu\electronvolt}$. At more positive gate voltages [Fig.~\ref{fig:3}(c)], where the carrier density is higher, peaks are 2{\em e}-periodic at zero field, then transition through even-odd to 1{\em e}-periodic Coulomb blockade without an overshoot, with an effective $g$-factor of $\sim 31$.

Overshoot of peak spacing, with $S_{\rm O}$ exceeding $S_{\rm E}$, indicates a discrete subgap state crossing zero energy \cite{Albrecht2016, Ciu2017}, consistent with interacting Majorana modes. The overshoot observed at more negative $V_{\rm G}$ is quantitatively in agreement with the overshoot seen in VLS wires of comparable length \cite{Albrecht2016}. The absence of the overshoot and the increase of the $g$-factor at positive $V_{\rm G}$ is consistent with the gate-tunable carrier density in VLS wires \cite{Vaitiekenas2017}.

\begin{figure}[t]
\includegraphics[width=\linewidth]{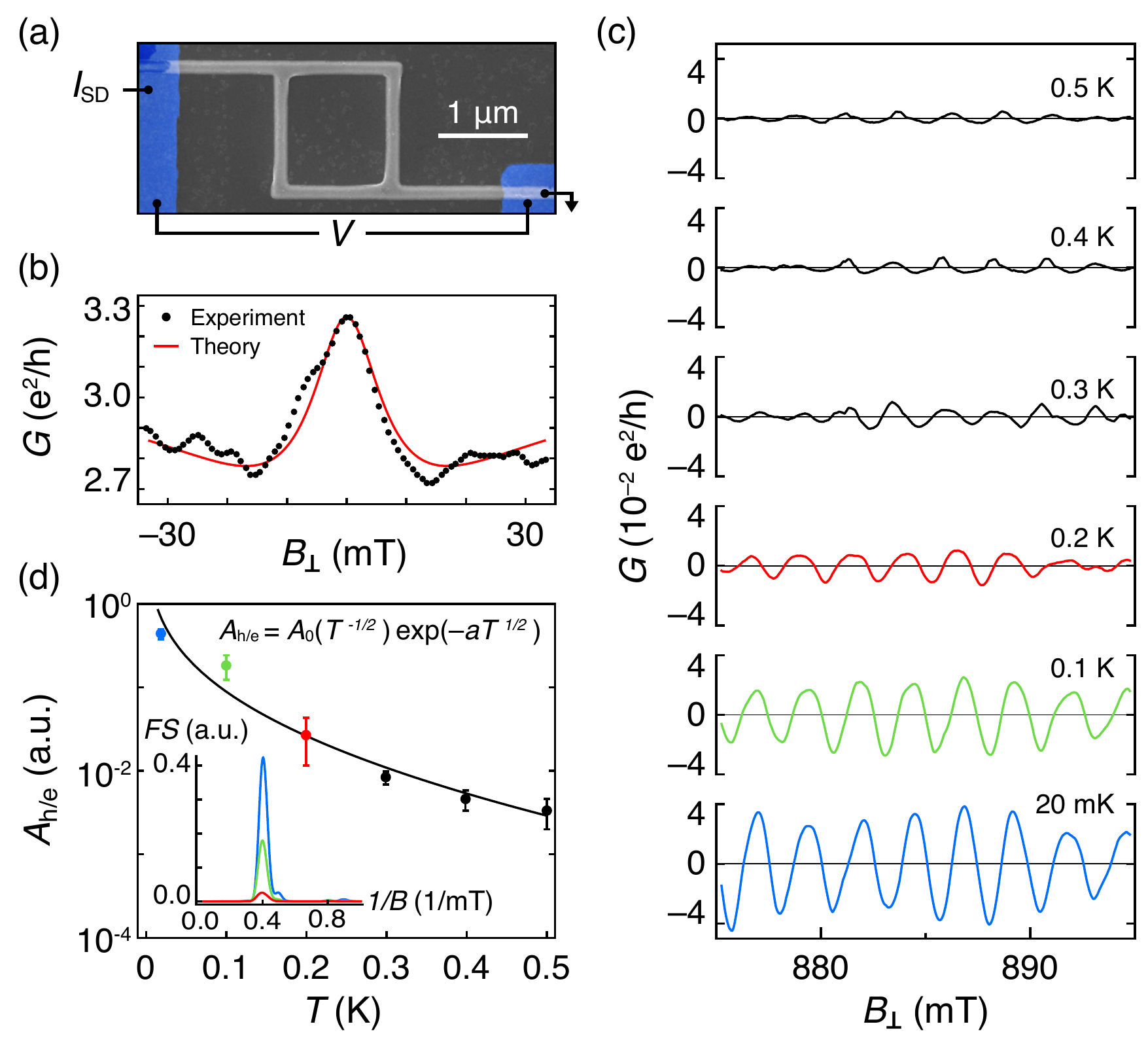}
\caption{\label{fig:4} (a) Electron micrograph of Device 3 with false-colored epitaxial Al contacts (blue) and overlaid 4-wire measurement setup. (b) Conductance, $G$, as a function $B_\perp$. Red curve is a theoretical magnetoconductance displaying the weak antilocalization effect in a system with spin-orbit length, $l_{\rm SO}\sim\SI{0.4}{\mu\meter}$, and phase-coherence length, $l_\phi^{\rm WAL}\sim\SI{1.2}{\mu\meter}$. (c) Aharonov-Bohm oscillations around $B_\perp\sim\SI{900}{\milli\tesla}$ at different temperatures. (d) Amplitude of the $h/e$ oscillations as a function of $T$. The exponential fit corresponds to the base-temperature phase-coherence length of $l_\phi^{\rm AB}=\SI{3.9}{\mu\meter}$. The error-bars correspond to the standard deviation between 4 data sets. Inset: Fourier spectra, $FS$, of the interference signal at $\SI{20}{\milli\kelvin}$ (blue), $\SI{100}{\milli\kelvin}$ (green) and $\SI{200}{\milli\kelvin}$ (red) from the measurements shown in (b).
}
\end{figure}

To demonstrate fabrication and operation of a simple SAG network, we investigate the coherence of electron transport in the loop structure shown in Fig.~\ref{fig:4}(a), with the Al layer completely removed by wet etching. Conductance as a function of perpendicular magnetic field, $B_\perp$, shows a peak around zero magnetic field, characteristic of WAL [Fig.~\ref{fig:4}(b)]. A fit to a theoretical model for disordered quasi one-dimensional wires with strong spin-orbit coupling \cite{Kurdak1992} yields an electron phase-coherence length $l_\phi^{\rm WAL}\sim\SI{1.2}{\mu\meter}$, and a spin-orbit length $l_{\rm SO}\sim\SI{0.4}{\mu\meter}$. We note that electrons propagating along $[100]$ and $[010]$ directions experience both Rashba and linear-Dresselhaus spin-orbit fields \cite{Sasaki2014}. The magnitude of each field can be deduced from a combination of in-plane magnetic field angle and magnitude dependence of the conductance correction due to the weak (anti-) localization. Such study, however, is out of scope of this work.

Upon suppressing WAL with a large perpendicular field periodic conductance oscillations are observed [Fig.~\ref{fig:4}(c)] with period $\Delta B_\perp = \SI{2.5}{\milli\tesla}$ corresponding to $h/e$ AB oscillations with area $\SI{1.7}{\mu\meter}^{2}$, matching the lithographic area of the loop. The oscillation amplitude, $A_{h/e}$, measured from the power spectral band around $h/e$ [Fig.~\ref{fig:4}(d), inset] was observed to decrease with increasing temperature as seen in Fig.~\ref{fig:4}(d). The size of $A_{h/e}$ is dictated by two characteristic lengths–thermal length $L_{\rm T}$ and phase-coherence length $l_\phi^{\rm AB}$ \cite{Kurdak1992}. We estimate $L_{\rm T}$ at $\SI{20}{\milli\kelvin}$ to be around $\SI{1.5}{\mu\meter}$ using a charge carrier mobility $\mu=\SI[per-mode=symbol]{700}{\centi\meter\squared\per\volt\per\second}$ and density $n=\SI{9e17}{\per\centi\meter\cubed}$ (see Supplemental Material \cite{SupMaterial}). The thermal length is comparable to the loop circumference $L=\SI{5.2}{\mu\meter}$, as a result, energy averaging is expected to have finite contribution to the size of the conductance oscillations \cite{Kurdak1992}. Taking $A_{h/e}\propto L_{\rm T}\left(T\right)\exp[-L/l_\phi^{\rm AB}\left(T\right)]$ with $L_{\rm T} \propto T^{-1/2}$ and $l_\phi^{\rm AB}\propto T^{-1/2}$ for a diffusive ring \cite{Ludwig2004}, a fit of the logarithmic amplitude $\log(A_{h/e}) = \log(A_0)-\frac{1}{2}\log{T}-aT^{1/2}$ yields $\log(A_0)\sim\SI{-1.5}{}$ and $a\sim\SI{6.7}{\kelvin\tothe{-1/2}}$ (Fig.~\ref{fig:4}d), giving a base-temperature phase-coherence length $l_\phi^{\rm AB}\left(\SI{20}{\milli\kelvin}\right)\sim\SI{5.5}{\mu\meter}$.

The discrepancy between the extracted $l_\phi^{\rm WAL}$ and $l_\phi^{\rm AB}$ has previously been observed in an experiment on GaAs/AlGaAs-based arrays of micron-sized loops \cite{Ferrier2008}. It has been argued theoretically that WAL and AB interference processes are governed by different dephasing mechanisms \cite{Ludwig2004}. As a result, $l_\phi^{\rm WAL}$ and $l_\phi^{\rm AB}$ have different temperature dependences.

Our results show that selective area grown hybrid nanowires are a promising platform for scalable Majorana networks exhibiting strong proximity effect. The hard induced superconducting gap and 2{\em e}-periodic Coulomb oscillations imply strongly suppressed quasiparticle poisoning. The overshoot of Coulomb peak spacing in a parallel magnetic field indicates the presence of a discrete low-energy state. Despite the relatively low charge carrier mobility, the measured SAG-based network exhibits strong spin-orbit coupling and phase-coherent transport. Furthermore, the ability to design hybrid wire planar structures containing many branches and loops---a requirement for realizing topological quantum information processing---is readily achievable in SAG. Future work on SAG-based hybrid networks will focus on spectroscopy, correlations, interferometry, and manipulation of MZMs.

We thank D. Carrad, A. Higginbotham and R. Lutchyn for valuable discussions as well as R. McNeil, C. S\o rensen and S. Upadhyay for contributions to material growth and device fabrication. The research was supported by Microsoft, the Danish National Research Foundation, and the European Commission. C.M.M.~acknowledges support from the Villum Foundation. M.T.D.~acknowledges support from State Key Laboratory of High Performance Computing, China. S.M-S. acknowledges funding from "Programa Internacional de Becas "la Caixa"-Severo Ochoa". ICN2 acknowledges support from the Severo Ochoa Programme (MINECO, Grant no. SEV-2013-0295) and is funded by the CERCA Programme / Generalitat de Catalunya. Part of the present work has been performed in the framework of Universitat Aut\`onoma de Barcelona Materials Science PhD program.

\onecolumngrid
\clearpage
\onecolumngrid
\setcounter{figure}{0}
\setcounter{equation}{0}
\section{\large{S\MakeLowercase{upplemental} M\MakeLowercase{aterial}
}}
\renewcommand{\figurename}{FIG.~S}
\twocolumngrid
\section{Sample Preparation}
The InAs nanowires with triangular cross-sections were selectively grown by MBE along the $[100]$ and $[010]$ directions on a semi-insulating $\left(001\right)$ InP substrate with a pre-patterned ($\SI{15}{\nano\meter}$) SiO$_{\rm x}$ mask \cite{supKrizek2017_2}. A thin ($\SI{7}{\nano\meter}$) layer of Al was grown \textit{in-situ} at low temperatures on one facet by angled deposition, forming an epitaxial interface with InAs. For the fabrication of the devices, Al was selectively removed using electron-beam lithography and wet etch (Transene Al Etchant D, \SI{50}{\celsius}, \SI{10}{\second}). Normal Ti/Al (\SI{5/120}{\nano\meter}) ohmic contacts were deposited after \textit{in-situ} Ar milling (RF ion source, \SI{15}{\watt}, \SI{18}{\milli\torr}, \SI{5}{\minute}). A film of HfO$_2$ ($\SI{7}{\nano\meter}$) was applied via atomic layer deposition at $\SI{90}{\celsius}$ before depositing Ti/Au (\SI{5/100}{\nano\meter}) gate electrodes.

\section{Measurement Setup}
The measurements were carried out with a lock-in amplifier at $f_{\rm ac}=\SI{173}{\hertz}$ in a dilution refrigerator with a base temperature of $T_{\rm base}=\SI{20}{\milli\kelvin}$. For voltage bias measurements, an ac signal with an amplitude of $\SI{0.1}{\volt}$ was applied to a sample through a homebuilt resistive voltage-divider $\left(1:17.700\right)$, resulting in $V_{\rm AC}\sim\SI{6}{\mu\volt}$ excitation. For current bias, we applied $\SI{2}{\volt}$ ac signal to a $\SI{1}{\giga\ohm}$ resistor in series with a sample giving $I_{\rm AC}=\SI{2}{\nano\ampere}$ excitation.

\section{Spectrum Evolution with $\bf V_{\bf\rm G}$}
\begin{figure}[b!]
\includegraphics[width=\linewidth]{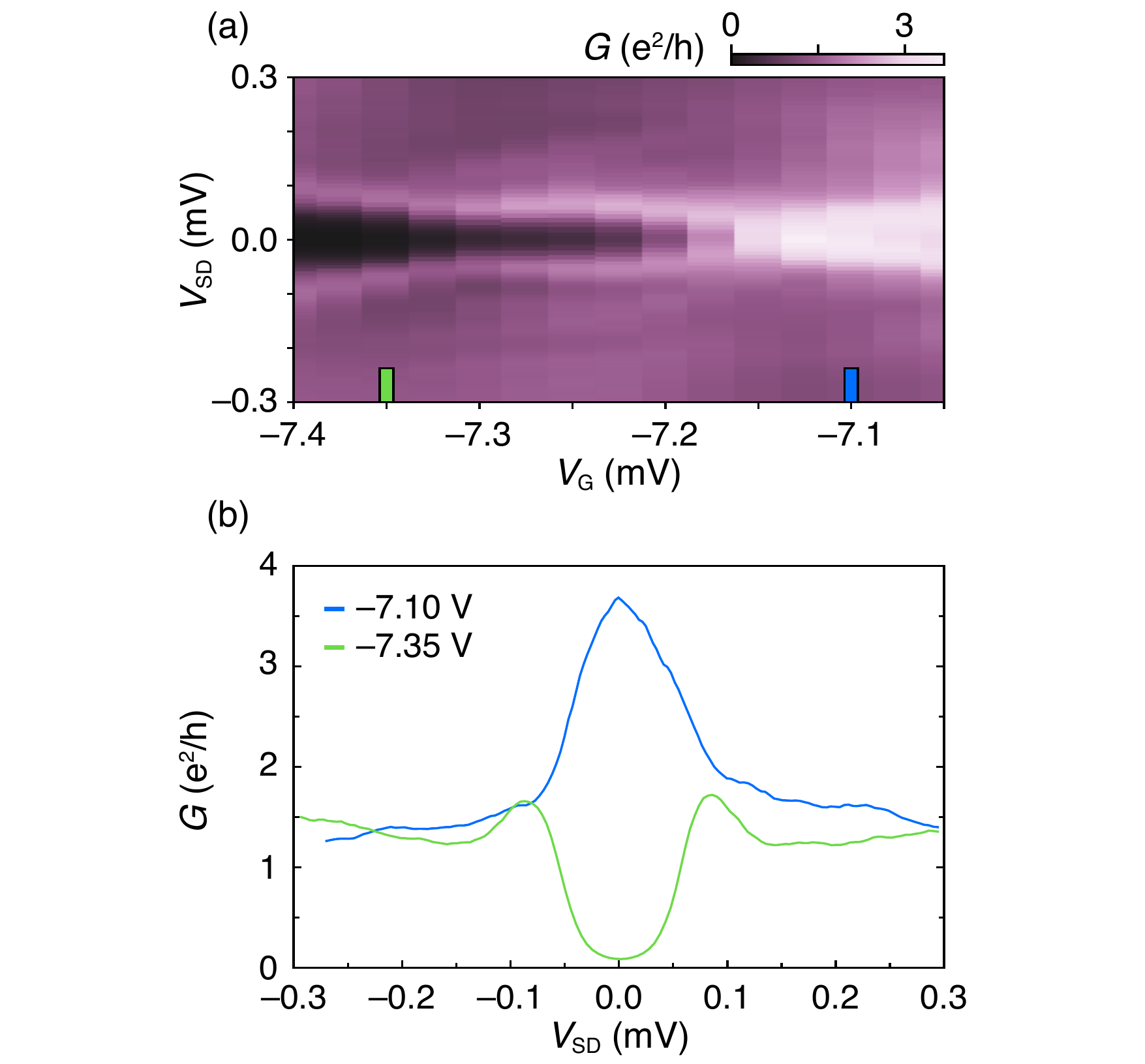}
\caption{\label{sfig:1} (a) Differential conductance, $G$, of Device 1 [Fig.~1(c,d), main text] as a function of source-drain bias, $V_{\rm SD}$, and gate voltage, $V_{\rm G}$. (b) Line-cuts taken from (a)
display Adreev reflection enhanced conductance (blue) measured at $V_{\rm G}=\SI{-7.10}{\volt}$ evolve into superconducting gap (green) measured at $V_{\rm G}=\SI{-7.35}{\volt}$.}
\end{figure}

Evolution of differential conductance, $G$, as a function of source-drain bias, $V_{\rm SD}$, with gate voltage, $V_{\rm G}$, for Device 1 [Fig.~1(c,d), main text] is illustrated in Fig.~S~\ref{sfig:1}(a). Conductance enhancement in the range of $V_{\rm SD}=\SI{-0.1}{\milli\volt}$ to $\SI{0.1}{\milli\volt}$ is measured at $V_{\rm G}=\SI{-7.10}{\volt}$ [Fig.~S~\ref{sfig:1}(b)]. At more negative gate voltage $V_{\rm G}=\SI{-7.35}{\volt}$ the conductance gets suppressed in the same range of $V_{\rm SD}$.

We interpret these features to arise due to a different tunnel barrier strengths tunned by the capacitively cross-coupled $V_{\rm G}$: At more positive $V_{\rm G}$, the tunneling barrier is more transparent, resulting in Andreev reflection enhanced conductance; At more negative $V_{\rm G}$, the transport is dominated by the single electron tunneling, reflecting the local density of states at the end of the wire \cite{supBlonder1982}. At $V_{\rm G} = \SI{-7.10}{\volt}$ the zero-bias conductance is enhanced beyond the factor of $2$ compared to high bias [Fig.~S~\ref{sfig:1}(b)]. We speculate that this is caused by an interfering transport mediated via multiple channels or a quantum dot \cite{supBeenakker1992}. A similar study of $V_{\rm C}$ dependence was not possible presumably due to a too high concentration of disorder in the junction.

\section{Free Energy Model}
\begin{figure}[t!]
\includegraphics[width=\linewidth]{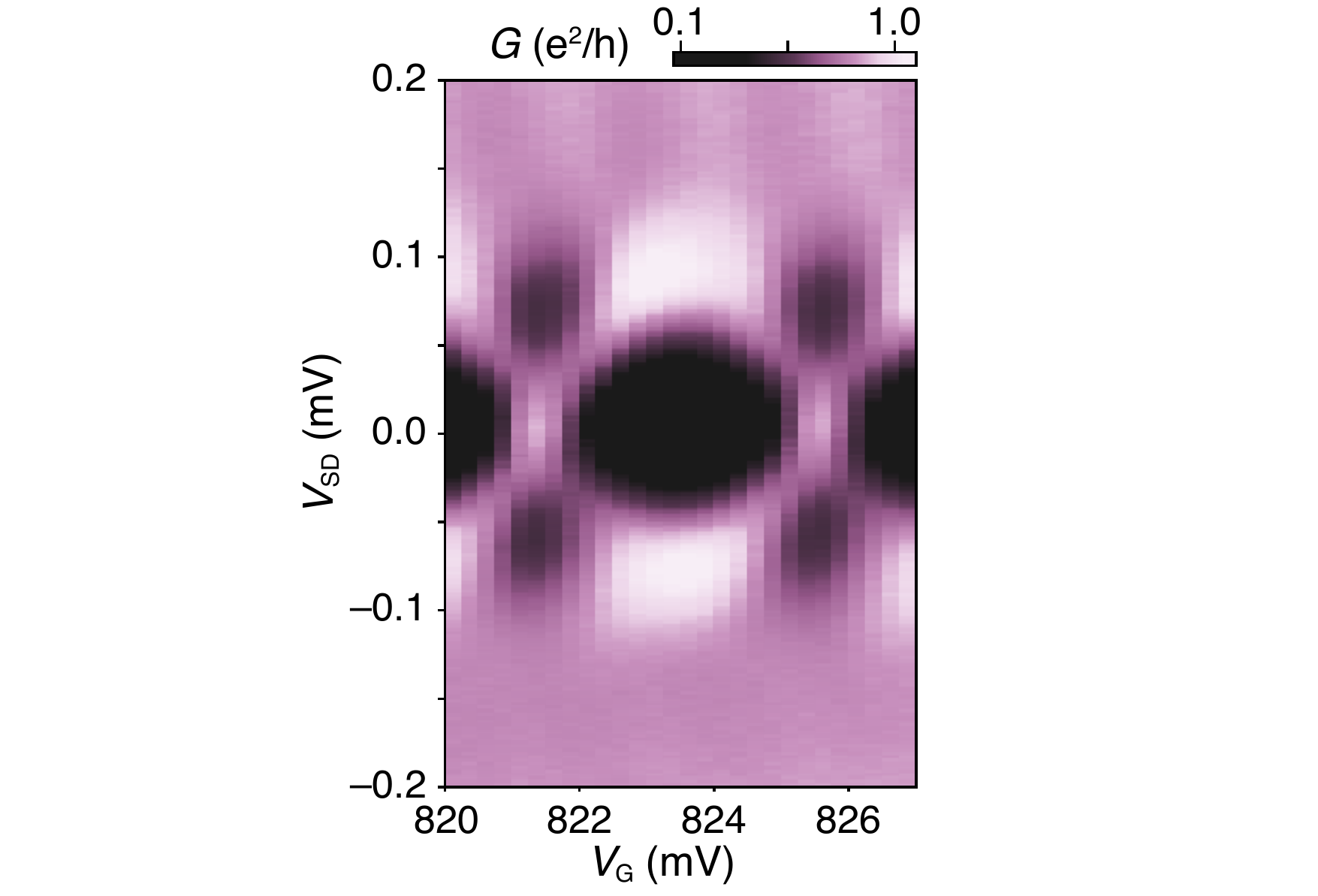}
\caption{\label{sfig:2} Differential conductance, $G$, of Device 2 as a function of source-drain bias, $V_{\rm SD}$, and gate voltage, $V_{\rm G}$ at zero magnetic field shows a Coulomb diamond with $2E_{\rm C}=\SI{120}{\mu\electronvolt}$.}
\end{figure}

The theoretical fit in Fig. 2(b) of the main text is based on a free energy model given by Eq. 1 in the main text, where the difference in free energy between odd and even occupied states is given by
\begin{equation}
\tag{S1}
\begin{split}
&F (T) =k_{\rm B}T\\&\times\ln\left[\frac{
\left(1+e^{-\Delta^*/k_{\rm B}T}\right)^{N_{\rm eff}}+
\left(1-e^{-\Delta^*/k_{\rm B}T}\right)^{N_{\rm eff}}
}{
\left(1+e^{-\Delta^*/k_{\rm B}T}\right)^{N_{\rm eff}}-
\left(1-e^{-\Delta^*/k_{\rm B}T}\right)^{N_{\rm eff}}
}\right],
\end{split}
\end{equation}
\noindent with the effective number of continuum states $N_{\rm eff}=2V_{\rm Al}\rho_{\rm Al}\,\sqrt[]{2\Delta^* k_{\rm B}T}$, where $V_{\rm Al}$ is the volume of the island and $\rho_{\rm Al}$ is the density of states at the Fermi energy \cite{supTuominen1992}. The fit was obtained by using $V_{\rm Al}=\SI{2.2e-6}{\cubic\nano\meter}$, consistent with Fig.~2(a) in the main text, electron density of states $\rho_{Al}=\SI{23}{\per\electronvolt\per\cubic\nano\meter}$ \cite{supTuominen1992} and $E_{\rm C}=\SI{60}{\milli\electronvolt}$, measured from Coulomb diamonds [Fig.~S~\ref{sfig:2}], with $\Delta^*$ as the single fit parameter.

\section{Device Energy}
Energy, $E$, of a Coulomb blockaded device with electron occupancy, $n$, as a function of normalized gate voltage, $N$, can be defined as
\begin{equation}\label{seq:2}
\tag{S2}
E(N) = E_{\rm C} (n-N)^2 + F N_0,
\end{equation}
where $E_{\rm C}$ is the charging energy, $F$ is the relative free energy and $N_0=0$ $(1)$ for even (odd) parity of the device, see Fig. 2(c) in the main text, inset. Charge degeneracy points can be extracted using Eq.~(\ref{seq:2}), from which we can deduce the normalized even and odd peak spacings in units of charge, $e$, as
\begin{equation}
\tag{S3}
N_{\rm E,O}=1 \pm F/E_{\rm C}.
\end{equation}
The even and odd peak spacing difference in gate voltage is given by
\begin{equation}
\tag{S4}
S_{\rm E}-S_{\rm O} = \frac{E_{\rm C}}{e \eta} (N_{\rm E}-N_{\rm O}),
\end{equation} with the dimensionless lever arm $\eta=E_{\rm C}/eS$.

Note that in the limit of zero temperature, $F$ is defined by the size of the induced gap, $\Delta^*$, or, if present, by the energy of a subgap state, $E_0$.

\section{Field Effect Mobility}
Conductance of a nanowire channel as a function of gate voltage is given by
\begin{equation}
\tag{S5}
G=\frac{\mu C}{l^2}\left(V_{\rm G}-V_{\rm T}\right),
\end{equation} where $\mu$ is the mobility, $C$ is the capacitance between the gate electrode and the wire, $l$ is the length of the channel, $V_{\rm G}$ is the gate voltage and $V_{\rm T}$ is the threshold voltage \cite{supGul2015}. By introducing transconductance, ${\rm d}G/{\rm d}V_{\rm G}$, the mobility can be expressed by
\begin{equation}\label{eq:6}
\tag{S6}
\mu=\frac{l^2}{C}\frac{{\rm d}G}{{\rm d}V_{\rm G}}.
\end{equation} 

\begin{figure}[t!]
\includegraphics[width=\linewidth]{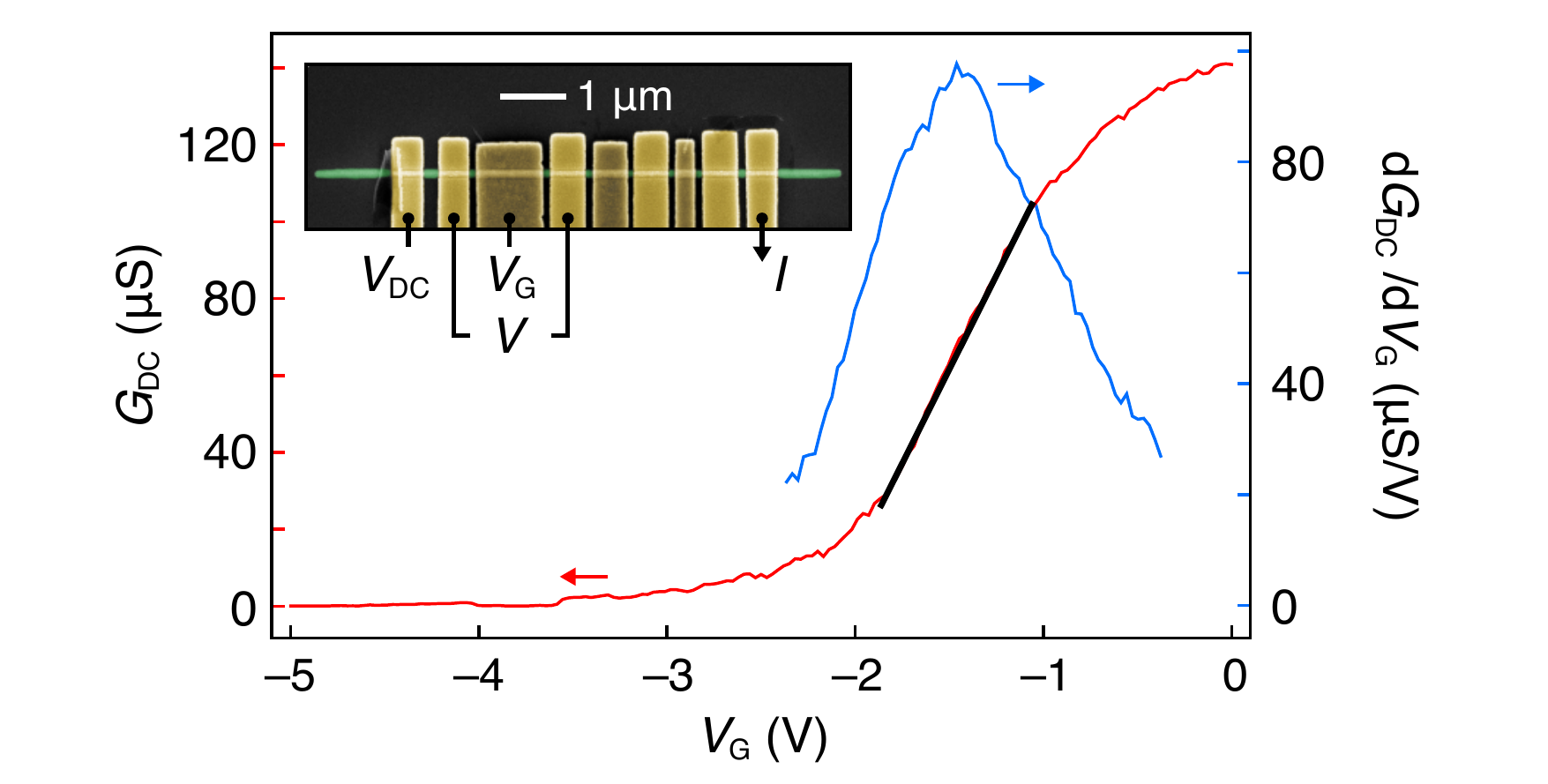}
\caption{\label{sfig:3} Conductance, $G_{\rm DC}$, (left axis) and transconductance, ${\rm d}G_{\rm DC}/{\rm d}V_{\rm G}$, (right axis) of Device 4 as a function of gate voltage, $V_{\rm G}$. The steepest slope extracted from the transconductance---indicated by the black line---yields field effect mobility $\mu=\SI[per-mode=symbol]{700}{\centi\meter\squared\per\volt\per\second}$. Inset: False-colored electron micrograph of Device 4 overlaid with the measurement setup.}
\end{figure}

Conductance, $G_{\rm DC}=I/V$, of Device 4 [Fig.~S~\ref{sfig:3} inset] measured at $\sim\SI{4}{\kelvin}$ and $V_{\rm DC}=\SI{5}{\milli\volt}$ as a function of top-gate voltage, $V_{\rm G}$, is shown in Fig.~S~\ref{sfig:3}. The transconductance peaks to $\sim\SI[per-mode=symbol]{100}{\mu\siemens\per\volt}$ around $V_{\rm G}=\SI{-1.5}{\volt}$, corresponding to the highest change in conductance indicated by the black line. The nanowire length $l=\SI{1}{\mu\meter}$ is set by the distance between the contacts. The capacitance $C=\SI{1.5}{\femto\farad}$ was estimated using COMSOL modeling software. Using Eq.~\ref{eq:6} results in $\mu=\SI[per-mode=symbol]{700}{\centi\meter\squared\per\volt\per\second}$.

Measurements on a similar chemical beam epitaxy grown SAG Hall bar with comparable mobility result in charge carrier density of $n\sim\SI{9e17}{\per\centi\meter\cubed}$ \cite{supCasparisUnpublished}. The corresponding mean free path is
\begin{equation}
\tag{S7}
l_{\rm e}=\frac{\hbar\mu}{e}\left(3\pi^2n\right)^{1/3}\sim\SI{15}{\nano\meter},
\end{equation} where $\hbar$ is the reduced Planck constant and $e$ is the elementary charge.

\section{Thermal Length}
The size of the Aharonov-Bohm oscillations is dictated by two characteristic length scales, namely the phase-coherence and thermal lengths \cite{supKurdak1992}. The thermal length is related to the energy averaging of conduction channels due to the finite electron temperature and is given by
\begin{equation}
\tag{S8}
L_{\rm T}\left(T\right)=\sqrt[]{\hbar D/k_{\rm B} T},
\end{equation} where $D$ is the diffusion constant and $k_{\rm B}$ is the Boltzmann constant. The diffusion constant is given by
\begin{equation}
\tag{S9}
D=\frac{1}{3}v_{\rm F}l_{\rm e},
\end{equation} where $v_{\rm F}=\hbar k_{\rm F}/m^*$ is the Fermi velocity, with the Fermi wave vector $k_{\rm F}=(3 \pi^2 n)^{1/3}$ and the effective electron mass in InAs $m^*=0.026 m_{\rm e}$, yielding $D=\SI[per-mode=symbol]{0.006}{\meter\squared\per\second}$, consistent with the values measured in vapor-liquid-solid (VLS) nanowires \cite{supJespersen2009}. The resulting thermal length $L_{\rm T}\left(\SI{20}{\milli\kelvin}\right)=\SI{1.5}{\mu\meter}$ is comparable to the loop circumference $L=\SI{5.2}{\mu\meter}$ in Device 3 [Fig. 4(a), main text].


\end{document}